\documentclass[pre, twocolumn, showkeys, amsmath, amssymb]{revtex4}

\usepackage{graphicx}

\begin{document}

\title{Sexual replication in the quasispecies model}

\author{Emmanuel Tannenbaum}
\email{emanuelt@bgu.ac.il}
\affiliation{Department of Chemistry, Ben-Gurion University of the Negev,
Be'er-Sheva 84105, Israel}

\begin{abstract}

This paper develops a simplified model for sexual replication within the quasispecies
formalism.  We assume that the genomes of the replicating organisms are two-chromosomed and 
diploid, and that the fitness is determined by the number of chromosomes that are identical
to a given master sequence.  We also assume that there is a cost to sexual replication,
given by a characteristic time $ \tau_{seek} $ during which haploid cells seek out a mate
with which to recombine.  If the mating strategy is such that only viable haploids can mate,
then when $ \tau_{seek} = 0 $, it is possible to show that sexual replication will always
outcompete asexual replication.  However, as $ \tau_{seek} $ increases, sexual replication
only becomes advantageous at progressively higher mutation rates.  Once the time cost for
sex reaches a critical threshold, the selective advantage for sexual replication disappears
entirely.  The results of this paper suggest that sexual replication is not advantageous
in small populations per se, but rather in populations with low replication rates.  In this
regime, the cost for sex is sufficiently low that the selective advantage obtained through
recombination leads to the dominance of the strategy.  In fact, at a given replication rate and
for a fixed environment volume, sexual replication is selected for in high populations 
because of the reduced time spent finding a reproductive partner.

\end{abstract}

\keywords{Sexual replication, diploid, haploid, quasispecies, error catastrophe,
error cascade, survival of the flattest}

\maketitle
                                                  
\section{Introduction}

Sexual reproduction is the observed mode of reproduction for nearly all multicellular
organisms.  As such, the evolution of sex has been one of the central outstanding questions
in evolutionary biology.

One of the biological explanations for the existence of sex is that it provides a natural
mechanism for diploid organisms to eliminate deleterious mutations from a population
\cite{SEX1}.  The idea is that, by reproducing via a haploid intermediate, it is possible for 
haploids without defective genes to recombine with one another, thereby preventing
the accumulation of deleterious mutations.  Other explanations that have been advanced
are that sex leads to greater variability in a population, making the population more adaptable
in adverse conditions.  It has also been postulated that sex evolved as a mechanism for
coping with parasites \cite{SEX1}.

In recent years, there have been a number of numerical studies focusing on the evolutionary
dynamics of sexual replication \cite{SEX1, SEX2, SEX3, SEX4, SEX5}.  These studies
have established that, depending on the choice of parameters, either sexual
or asexual modes of reproduction are the advantageous replication strategy.  One study
in particular argues that sexual reproduction is favored when the
number of daughter genomes produced by the parents is high, since this reduces the
amount of time required to find a reproductive partner \cite{SEX3}.

In this paper, we present a relatively simple evolutionary dynamics model that allows
us to compare sexual and asexual replication strategies.  The model is analytically
solvable, and treatable within the quasispecies formalism \cite{EIG}.  The essential result
is that sexual replication is favored in populations with low replication rates,
and when the characteristic time associated with finding a reproductive partner is
small compared with the time scale associated with replication.  These results 
suggest that increasing population density favors the sexual replication strategy,
since it reduces the time scale associated with finding a mate. 

This paper is organized as follows:  In the following section (Section II), we develop
a simplified model for sexual replication, whose steady-state behavior we proceed
to characterize in Section III.  In Section IV we compare sexual and asexual replication,
and establish regimes where each is the preferred mode of reproduction.  We conclude
the paper in Section V with a brief discussion and an outline of avenues for future
research.

\section{A Simplified Model for Sexual Replication}

In a simplified model for sexual replication, we assume that we have a population
of single-celled organisms, where each organism has a genome consisting of
two chromosomes.  We assume that each chromosome may be denoted by a linear
symbol sequence $ \sigma = s_1 \dots s_L $, where each letter, or base, $ s_i $,
is chosen from an alphabet of size $ S $ ($ = 4 $ for terrestrial life).  We
further assume that there exists a ``master'' sequence $ \sigma_0 $ for which
a given chromosome is functional.  It is assumed that a chromosome is nonfunctional
whenever $ \sigma \neq \sigma_0 $ (that is, the genes on such a chromosome are
defective).

Within this approximation, there are three distinct types of genomes in the
population:  (1) $ \{\sigma_0, \sigma_0\} $ -- Genomes where both chromosomes
are identical to the ``master'' sequence.  (2) $ \{\sigma_0, \sigma \neq \sigma_0\} $ --
Genomes where only one of the chromosomes is identical to the ``master'' sequence,
while the other chromosome is defective.  (3) $ \{\sigma \neq \sigma_0,
\sigma' \neq \sigma_0\} $ -- Genomes where both of the chromosomes are defective.

We are therefore dealing with a diploid population.  If we assign the gene
sequence $ \sigma_0 $ as {\it Viable}, while all other gene sequences are
{\it Unviable}, then our three genome types may be classified as
$ \{V, V\} $, $ \{V, U\} $, and $ \{U, U\} $, where $ V/U $ stand for
Viable/Unviable.  

We assume that the organisms replicate with a first-order growth rate constant.
For the three distinct genome types, the first-order growth rate constants
are taken to be $ \kappa_{VV} $, $ \kappa_{VU} $, and $ \kappa_{UU} $.
We have that $ \kappa_{VV} \geq \kappa_{VU} > \kappa_{UU} $.

The sexual replication of the population occurs as follows:  The diploid
organisms divide to form a population of haploid organisms.  It is assumed that
those haploid organisms containing a genome of type U are incapable of
participating further in the reproductive process, so that only viable haploids
can recombine with each other.  The newly formed diploids then divide via the
normal mitotic pathways to form two new daughter cells.

To develop a set of ordinary differential equations governing the
replication dynamics described above, we first let $ n_{VV} $ denote
the total number of organisms with genome of type $ \{V, V\} $,
$ n_{VU} $ denote the total number of organisms with genome of type
$ \{V, U\} $, and $ n_{UU} $ denote the total number of organisms with
genome of type $ \{U, U\} $.  We also let $ n_V $ denote the population
of viable haploids.  We then wish to obtain expressions for $ d n_{VV}/dt $,
$ d n_{VU}/dt $, $ d n_{UU}/dt $, and $ d n_V/dt $.  

First note that, the diploid to haploid division leads to destruction of 
each of the diploid genomes at a rate given by $ -\kappa_{VV} n_{VV} $
for $ \{V, V\} $, and similarly for the other genomes, and a creation of
viable haploid genomes at a rate given by $ 2 \kappa_{VV} n_{VV} +
\kappa_{VU} n_{VU} $.  

If we let $ \tau_{seek} $ denote the average amount of time a viable haploid spends
searching for a viable haploid mate, then in a given amount of time $ dt $
the total number of viable haploids who have recombined is given by 
$ n_V dt/\tau_{seek} $ (the individual times are Poisson distributed).  Therefore, 
recombination leads to a destruction rate of haploids given by $ n_V/\tau_{seek} $, 
and a creation rate of diploids given by $ (1/2) n_V/\tau_{seek} $.  

If we let $ p $ denote the probability of correctly replicating a chromosome,
then neglecting backmutations we have that $ V \rightarrow V $ with probability
$ p $, $ V \rightarrow U $ with probability $ 1 - p $, and $ U \rightarrow U $
with probability $ 1 $.  Using this, we can construct the various possible
replication pathways and their associated probabilities, illustrated in
Figure 2.  From these pathways, we can construct the contribution to
$ n_{VV} $, $ n_{VU} $, and $ n_{UU} $ in turn.

\begin{figure}
\includegraphics[width = 0.9\linewidth, angle = 0.0]{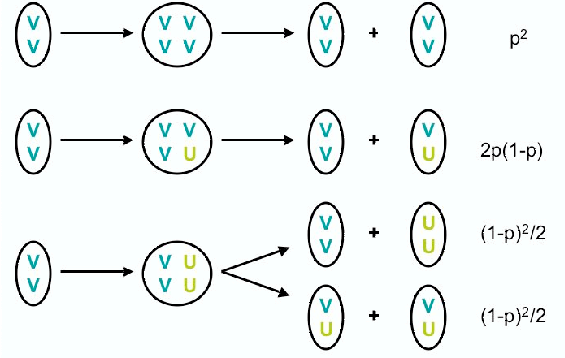}
\caption{The various replication pathways and their associated probabilities.
The factor of $ 2 $ in the second pathway comes from the fact that either the top
or bottom parent ``V'' chromosome can form a daughter ``U'' chromosome.}
\end{figure}

For $ n_{VV} $, all three replication pathways in Figure 1 give a contribution.
Taking into account probabilities and degeneracies, we have a total contribution
of $ 2 p^2 $ from the first pathway, $ 2 p(1-p) $ from the second pathway,
and $ (1-p)^2/2 $ from the second pathway, giving a total rate of production
of $ (1/2) n_V/\tau_{seek} \times [2 p^2 + 2 p (1 - p) + (1 - p)^2/2]
= (1/4) n_V/\tau_{seek} (1 + p)^2 $. 

For $ n_{VU} $, similar reasoning gives a rate of production of
$ (1/2) n_V/\tau_{seek} \times [2 p (1 - p) + 2 (1 - p)^2/2]
= (1/2) n_V\tau_{seek} (1 - p^2) $.

For $ n_{UU} $, we obtain $ (1/2) n_V/\tau_{seek} \times
(1 - p)^2/2 = (1/4) n_V/\tau_{seek} (1 - p)^2 $.

Putting everything together, we obtain the system of differential
equations,
\begin{eqnarray}
&   &
\frac{d n_{VV}}{dt} = -\kappa_{VV} n_{VV} +
\frac{n_V}{4 \tau_{seek}} (1 + p)^2
\nonumber \\
&   &
\frac{d n_{VU}}{dt} = -\kappa_{VU} n_{VU} +
\frac{n_V}{2 \tau_{seek}} (1 - p^2)
\nonumber \\
&   &
\frac{d n_{UU}}{dt} = -\kappa_{UU} n_{UU} + \frac{n_V}{4 \tau_{seek}} (1 - p)^2
\nonumber \\
&   &
\frac{d n_V}{dt} = 2 \kappa_{VV} n_{VV} + \kappa_{VU} n_{VU} - \frac{n_V}{\tau_{seek}}
\end{eqnarray}

We can non-dimensionalize these equations by defining $ \tau = t/\tau_{seek} $,
and $ \tilde{\kappa}_{VV} = \kappa_{VV} \tau_{seek} $, $ \tilde{\kappa}_{VU}
= \kappa_{VU} \tau_{seek} $, $ \tilde{\kappa}_{UU} = \kappa_{UU} \tau_{seek} $.
Then we obtain,
\begin{eqnarray}
&   &
\frac{d n_{VV}}{d \tau} = -\tilde{\kappa}_{VV} n_{VV} +
\frac{n_V}{4} (1 + p)^2 
\nonumber \\
&   &
\frac{d n_{VU}}{d \tau} = -\tilde{\kappa}_{VU} n_{VU} +
\frac{n_V}{2} (1 - p^2)
\nonumber \\
&   &
\frac{d n_V}{dt} = 2 \tilde{\kappa}_{VV} n_{VV} + \tilde{\kappa}_{VU} n_{VU} -
n_V
\end{eqnarray}

Define $ n = n_{VV} + n_{VU} + n_{UU} + n_V $, $ n' = n_{VV} + n_{VU} + n_{UU} $,
$ x_{VV} = n_{VV}/n $, $ x_{VV}' = n_{VV}/n' $, $ x_{VU} = n_{VU}/n $, $ x_{VU}' =
n_{VU}/n' $, $ x_{UU} = n_{UU}/n $, $ x_{UU}' = n_{UU}/n' $, $ x_V = n_V/n $.  Note
that $ x_{VV}' = x_{VV}/(1 - x_V) $, $ x_{VU}' = x_{VU}/(1 - x_V) $, $ x_{UU}' =
x_{UU}/(1 - x_V) $, and $ n_V/n' = x_V/(1 - x_V) $.

Then defining $ \bar{\kappa}(\tau) = (1/n) dn/d \tau $, $ \bar{\kappa}(\tau)' = (1/n') dn'/d \tau $,
we have,
\begin{eqnarray}
&   &
\bar{\kappa}(\tau) = \tilde{\kappa}_{VV} x_{VV} - \tilde{\kappa}_{UU} x_{UU} 
\nonumber \\
&   &
\bar{\kappa}(\tau)' = -(\tilde{\kappa}_{VV} x_{VV}' + \tilde{\kappa}_{VU} x_{VU}' + \tilde{\kappa}_{UU} x_{UU}') 
\nonumber \\
&   &
+ \frac{x_V}{1 - x_V}
\end{eqnarray}

Re-expressing the dynamical equations in terms of the $ x_{VV} $, $ x_{VU} $, $ x_{UU} $, $ x_V $ population fractions,
we have,
\begin{eqnarray}
&   &
\frac{d x_{VV}}{d \tau} = -(\tilde{\kappa}_{VV} + \bar{\kappa}(\tau)) x_{VV} + \frac{x_V}{4} (1 + p)^2 
\nonumber \\
&   &
\frac{d x_{VU}}{d \tau} = -(\tilde{\kappa}_{VU} + \bar{\kappa}(\tau)) x_{VU} + \frac{x_V}{2} (1 - p^2)
\nonumber \\
&   &
\frac{d x_{UU}}{d \tau} = -(\tilde{\kappa}_{UU} + \bar{\kappa}(\tau)) x_{UU} + \frac{x_V}{4} (1 - p)^2
\nonumber \\
&   &
\frac{d x_V}{d \tau} = -(1 + \bar{\kappa}(\tau)) x_V + 2 \tilde{\kappa}_{VV} x_{VV} + \tilde{\kappa}_{VU} x_{VU}
\nonumber \\
\end{eqnarray} 

\section{Equilibrium Mean Fitness Results}

At steady-state, the above time derivatives may all be set to $ 0 $, giving,
\begin{equation}
x_V = \frac{2 \tilde{\kappa}_{VV} x_{VV} + \tilde{\kappa}_{VU} x_{VU}}{1 + \bar{\kappa}(\tau = \infty)}
\end{equation}
Therefore, at steady-state, we have that,
\begin{eqnarray}
\bar{\kappa}(\tau = \infty)' 
& = & 
-\frac{1}{1 - x_V}[((2 \tilde{\kappa}_{VV} x_{VV} + \tilde{\kappa}_{VU} x_{VU}) 
\nonumber \\
&   &
- (\tilde{\kappa}_{VV} x_{VV} - \tilde{\kappa}_{UU} x_{UU})] + \frac{x_V}{1 - x_V}
\nonumber \\
& = &
\frac{1}{1 - x_V}[x_V + \bar{\kappa}(\tau = \infty) - x_V (1 + \bar{\kappa}(\tau = \infty))]
\nonumber \\
& = &
\bar{\kappa}(\tau = \infty)
\end{eqnarray}
and so it is equivalent to measure the mean fitness of the population at steady-state using
either $ \bar{\kappa}(\tau) $ and $ \bar{\kappa}(\tau)' $.  This is important, because, in comparing
mean fitness results with an asexually replicating population, the natural mean fitness result
to use is $ \bar{\kappa}(\tau)' $.  The equivalence between $ \bar{\kappa}(\tau) $ and $ \bar{\kappa}(\tau)' $
means that we can compute $ \bar{\kappa}(\tau) $ at steady-state and compare the results directly
with the value of $ \bar{\kappa}(\tau) $ for the asexually replicating population.

Plugging the steady-state value of $ x_V $ into the steady-state equations for $ x_{VV} $ 
and $ x_{VU} $, we obtain,
\begin{eqnarray}
&   &
0 = [\tilde{\kappa}_{VV}(2(\frac{1+p}{2})^2 \frac{1}{1 + \bar{\kappa}(\tau = \infty)} - 1) - \bar{\kappa}(\tau = \infty)] x_{VV} 
\nonumber \\
&   &
+ \frac{1}{4} (1 + p)^2 \frac{\tilde{\kappa}_{VU} x_{VU}}{1 + \bar{\kappa}(\tau = \infty)}
\nonumber \\
&   &
0 = [\tilde{\kappa}_{VU} (\frac{1}{2} (1 - p^2) \frac{1}{1 + \bar{\kappa}(\tau = \infty)} - 1) - \bar{\kappa}(\tau = \infty)] x_{VU}
\nonumber \\
&   &
+ (1 - p^2) \frac{1}{1 + \bar{\kappa}(\tau)} \tilde{\kappa}_{VV} x_{VV}
\end{eqnarray}
We can solve the first equation for $ x_{VU} $ in terms of $ x_{VV} $.  Plugging the resulting expression
into the second equation, we obtain, after some algebra, the quadratic,
\begin{eqnarray}
0 
& = & 
\bar{\kappa}(\tau = \infty)^2 
\nonumber \\
&   &
- [\tilde{\kappa}_{VV}' (2 (\frac{1 + p}{2})^2 - 1) - \frac{1}{2} \tilde{\kappa}_{VU}' (1 + p^2)] \bar{\kappa}(\tau = \infty)
\nonumber \\
&   &
- \tilde{\kappa}_{VV}' \tilde{\kappa}_{VU}' p
\end{eqnarray}
where $ \tilde{\kappa}_{VV}' \equiv \tilde{\kappa}_{VV}/(1 + \tilde{\kappa}_{VV}) $, $ \tilde{\kappa}_{VU}' \equiv
\tilde{\kappa}_{VU}/(1 + \tilde{\kappa}_{VU}) $.

We can further simplify the notation by defining $ \kappa = \tilde{\kappa}_{VV} $, and
$ \alpha = \tilde{\kappa}_{VU}/\tilde{\kappa}_{VV} $.  Then $ \bar{\kappa}(\tau = \infty)/\kappa $ is
the solution to the quadratic,
\begin{equation}
0 = x^2 - A(p, \kappa, \alpha) x - B(p, \kappa, \alpha)
\end{equation}
where,
\begin{eqnarray}
&   &
A(p, \kappa, \alpha) = \frac{1}{1 + \kappa} [2 (\frac{1 + p}{2})^2 - 1] - \frac{1}{2} \frac{\alpha}{1 + \alpha \kappa} (1 + p^2)
\nonumber \\
&   &
B(p, \kappa, \alpha) = \frac{1}{1 + \kappa} \frac{\alpha}{1 + \alpha \kappa} p
\end{eqnarray}

Differentiating both sides of the quadratic, it is possible to show, after some
manipulation, that $ d x/d p > 0 $, and hence that the mean fitness is an
increasing function of $ p $.

\section{Comparison of Sexual and Asexual Replication}

If, for simplicity, we assume that $ \kappa_{UU} = 0 $, then for asexual replication the
steady-state value for $ \bar{\kappa}(\tau = \infty)/\kappa $ may be readily characterized
\cite{DIP}:  It is given by $ \max\{2 (\frac{1 + p}{2})^2 - 1, \alpha p\} $.
Therefore, if $ p_{crit} $ is defined by the equality $ 2 ((1 + p_{crit})/2)^2 - 1 = 
\alpha p_{crit} $, then,
\begin{eqnarray}
\frac{\bar{\kappa}(\tau = \infty)}{\kappa} =
\left\{\begin{array}{cc}
  2 (\frac{1 + p}{2})^2 - 1 & \mbox{if $ p \in [p_{crit}, 1] $} \\
  \alpha p & \mbox{if $ p \in (0, p_{crit}] $}
  \end{array}
\right.
\end{eqnarray}

We can compare these values with that obtained for sexual replication.

\subsection{Case 1:  $ \kappa = 0 $}

We begin by considering the case where there is no time cost associated with sex, 
so that $ \tau_{seek} = 0 \Rightarrow \kappa = 0 $.  Then
\begin{eqnarray}
&   &
A(p, 0, \alpha) = [2 (\frac{1 + p}{2})^2 - 1] - \frac{1}{2} \alpha (1 + p^2)
\nonumber \\
&   &
B(p, 0, \alpha) = \alpha p
\end{eqnarray}

We claim that for $ \kappa = 0 $, $ \bar{\kappa}(\tau = \infty)/\kappa 
\geq \max\{2 (\frac{1 + p}{2})^2 - 1, \alpha p\} $, with equality
occurring only when $ p = 1 $ for arbitrary $ \alpha $, and
$ \alpha = 0, 1 $.

To prove this claim, note first that for $ p = 1 $, we have
$ 2 (\frac{1 + p}{2})^2 - 1 = 1 $, and that
$ \bar{\kappa}(\tau = \infty)/\kappa = 1 = \max \{1, \alpha\} $.  So since
the claim is true for $ p = 1 $, we may now consider $ p \in [0, 1) $.

If $ \alpha = 0 $, then $ \bar{\kappa}(\tau = \infty) =
\max \{2 (\frac{1 + p}{2})^2 - 1, 0\} $, while if
$ \alpha = 1 $, then $ \bar{\kappa}(\tau = \infty) =
p = \max \{2 (\frac{1 + p}{2})^2 - 1, p\} $, since
$ 2 (\frac{1 + p}{2})^2 - 1 \leq p $ for $ p \in [0, 1] $,
with equality occuring only when $ p = 1 $.

So, we now consider the case where $ \alpha \in (0, 1) $,
and $ p \in [0, 1) $.  If we define $ k(p) = 
2 (\frac{1 + p}{2})^2 - 1 $, then we have two
possibilities:  Either $ \max \{k(p), \alpha p\} =
k(p) $, or $ \max\{k(p), \alpha p\} = \alpha p $.
We will consider each of these two cases in turn.

So, first assume that $ \max \{k(p), \alpha p\} = k(p) $.
We wish to show that,
\begin{equation}
\frac{1}{2}[k(p) - \frac{1}{2} \alpha (1 + p^2) +
\sqrt{(k(p) - \frac{1}{2} \alpha (1 + p^2))^2 + 4 \alpha p}] >
k(p)
\end{equation}
After some manipulation, we obtain that this condition is
equivalent to the condition that
\begin{equation}
0 > p^4 + 2 p^3 - 2 p - 1
\end{equation}
To establish this inequality for $ p \in [0, 1) $, note that 
$ p^4 + 2 p^3 - 2 p - 1 = (p - 1)(p^3 + 3 p^2 + 3 p + 1) $.
Since $ p^3 + 3 p^2 + 3 p + 1 > 0 $ for $ p \in [0, 1) $, and
since $ p - 1 < 0 $ for $ p \in [0, 1) $, the inequality
follows.

So now suppose that $ \max \{k(p), \alpha p\} = \alpha p $,
so that $ k(p) < \alpha p $.  Then for our calculations,
we first re-write $ A(p, 0, \alpha) $ 
as $ \alpha (p - 1) + (1 - \alpha) k(p) $.  Then we wish
to show that,
\begin{eqnarray}
&   &
\frac{1}{2}[\alpha (p - 1) + (1 - \alpha) k(p) +
\nonumber \\
&   &
\sqrt{(\alpha (p - 1) + (1 - \alpha) k(p))^2 + 4 \alpha p}]
\nonumber \\
&   &
> \alpha p
\end{eqnarray}
After some manipulation, this becomes equivalent to the condition
that,
\begin{eqnarray}
k(p) + 1 > \alpha (k(p) + 1)
\end{eqnarray}
which is certainly true, since $ k(p) + 1 > 0 $ for $ p \in [0, 1) $, and
$ \alpha \in (0, 1) $ by assumption.

Therefore, we have proven our claim, and hence, within this model, sexual 
replication leads to a greater mean fitness for a population than asexual
replication, assuming that there is no cost associated with sex.  Figure 2
shows a plot of $ \bar{\kappa}(\tau = \infty)/\kappa $ for both sexual
and asexual replication, assuming $ \alpha = 1/2 $ and $ \kappa = 0 $.

\begin{figure}
\includegraphics[width = 0.9\linewidth, angle = -90]{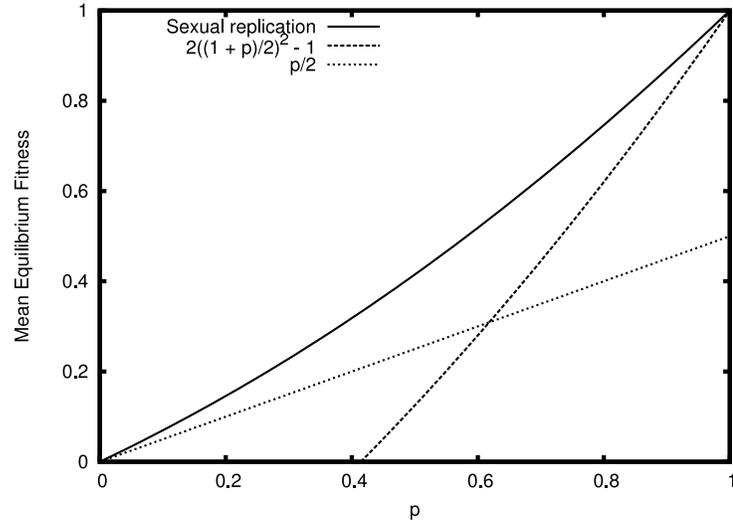}
\caption{Comparison of $ \bar{\kappa}(\tau = \infty)/\kappa $ for
both sexual and asexual replication, with $ \kappa = 0 $ and $ \alpha
= 1/2 $.  Note that sexual replication outcompetes asexual replication
for all mutation regimes.}
\end{figure}

Note that for $ \kappa = 0 $, if a sexually and asexually
replicating population were placed in an identical flask, then 
under the circumstances dictated by our model the sexually replicating
population would eventually dominate the population (that is, the fraction
of sexually replicating organisms would increase to $ 1 $, while the
fraction of asexually replicating organisms would decrease to $ 0 $). 

\subsection{Case 2:  $ \kappa > 0 $}

For $ \kappa > 0 $, we have, for sexual replication, that
$ \bar{\kappa}(\tau = \infty)/\kappa = \frac{1}{1 + \kappa} $ for $ p = 1 $.
Since for asexual replication we get $ \bar{\kappa}(\tau = \infty)/\kappa = 1 $
for $ p = 1 $, it follows by continuity that there exists a regime
$ [p_{=}(\kappa), 1] $ for which asexual reproduction leads to a greater
mean fitness of the population than sexual reproduction.  Presumably,
as $ \kappa $ increases, $ p_{=}(\kappa) $ should decrease.

We can determine, for a given $ \kappa $, the mutation regime where
asexual replication outcompetes sexual replication, and the
mutation regime where sexual replication outcompetes asexual
replication.  To do this, we do not attempt to compute $ p_{=}(\kappa) $
directly.  Rather, as a function of $ p $, we seek to determine
$ \kappa_{=}(p) $, the value of $ \kappa $ for which asexual and sexual replication
yield identical mean fitnesses.  Since by definition $ \kappa_{=}(p(\kappa)) = \kappa $,
the function $ \kappa_{=}(p) $ may be inverted to obtain $ p_{=}(\kappa) $.  In what
follows, we restrict our analysis to the case where $ \alpha \in (0, 1) $ (if $ \alpha = 0, 1 $,
then for $ \kappa > 0 $ asexual reproduction outcompetes sexual reproduction for all
values of $ p $).

Because the mean fitness of an asexually replicating population falls into two distinct
regimes defined by the cutoff $ p_{crit} $ (at least, within the context of our model), 
we must determine $ \kappa_{=}(p) $ separately for $ p \leq p_{crit} $ and $ p > p_{crit} $.
For $ p \leq p_{crit} $, we have for asexual replication that $ \bar{\kappa}(\tau = \infty)/\kappa 
= \alpha p $, and hence we must have,
\begin{eqnarray}
0 
& = &
(\alpha p)^2 - (\frac{1}{1 + \kappa_{=}} k(p) - \frac{1}{2} \frac{\alpha}{1 + \alpha \kappa_{=}} (1 + p^2)) (\alpha p)
\nonumber \\
&   &
- \frac{1}{1 + \kappa_{=}} \frac{1}{1 + \alpha \kappa_{=}} (\alpha p)
\end{eqnarray}
Assuming that $ \alpha p > 0 $, this expression may be re-arranged and simplified to,
\begin{equation}
0 = (\alpha \kappa_{=})^2 p + (\alpha p + 1) (\alpha \kappa_{=}) -
\frac{1}{2} (1 - \alpha) (1 + p)^2
\end{equation}
so that $ \alpha \kappa_{=} $, and hence $ \kappa_{=} $, may be
solved using the quadratic formula.

We claim that $ \kappa_{=} $ is an increasing function of $ p $
on the interval $ [0, p_{crit}] $.  We can prove this by
showing that $ \alpha \kappa_{=} $ is an increasing function
of $ p $ on the interval $ [0, p_{crit}] $.  Defining
$ x(p) = \alpha \kappa_{=}(p) $, we have,
\begin{equation}
0 = p x(p)^2 + (\alpha p + 1) x(p) - \frac{1}{2} (1 - \alpha) (1 + p)^2
\end{equation}
Differentiating with respect to $ p $, we obtain,
\begin{equation}
0 = (2 p x(p) + \alpha p + 1) x'(p) + x(p)^2 + \alpha x(p) - (1 - \alpha) (1 + p)
\end{equation}
so we wish to show that $ x(p)^2 + \alpha x(p) - (1 - \alpha) (1 + p) < 0 $ 
for $ p \in (0, p_{crit}) $.  Multiplying both sides of the inequality
by $ p $, and noting that $ p x(p)^2 = \frac{1}{2} (1 - \alpha) (1 + p)^2
- (\alpha p + 1) x(p) $, we have that for $ p > 0 $ we need to establish the inequality,
\begin{equation}
x(p) > \frac{1}{2} (1 - \alpha) (1 - p^2)
\end{equation}

To prove this inequality, note first that $ x(0) = \frac{1}{2} (1 - \alpha) $,
and $ x'(0) = \frac{1}{4} (1 - \alpha) (3 - \alpha) > 0 $.  By continuity,
$ x'(p) > 0 $ in a neighborhood of $ p = 0 $.  If $ x'(p) \leq 0 $ for
some $ p > 0 $, then by the Intermediate Value Theorem there exists
at least one $ p > 0 $ for which $ x'(p) = 0 $.  If 
$ p^{*} \equiv \inf \{p \in [0, 1]| x'(p) = 0\} $, then by
continuity it follows that $ x'(p^{*}) = 0 $, and hence
$ p^{*} > 0 $.  Therefore, $ x'(p) > 0 $ for $ p \in
[0, p^{*}) $, otherwise by the Intermediate Value Theorem
there would exist a $ p^{**} < p^{*} $ such that
$ x'(p^{**}) = 0 $, contradicting the definition
of $ p^{*} $.  But, since $ x'(p) > 0 $ for $ p \in
[0, p^{*}) $, it follows that $ x(p) $ is increasing
on $ [0, p^{*}] $, hence $ x(p^{*}) > x(0) =
\frac{1}{2} (1 - \alpha) > \frac{1}{2} (1 - \alpha) (1 - (p^{*})^2) $,
which implies that $ x'(p^{*}) > 0 \Rightarrow\Leftarrow $.  Therefore,
$ x'(p) > 0 $ for $ p \in [0, 1] $, hence on $ [0, p_{crit}] $,
$ \kappa_{=}(p) $ increases from $ \frac{1}{2} (1 - \alpha)/\alpha $
to $ \kappa_{=}(p_{crit}(\alpha)) $.

Now, for $ p \in [p_{crit}, 1] $, we have for an asexually
replicating population that $ \bar{\kappa}(\tau = \infty)/\kappa
= k(p) $, hence, in this regime, $ \kappa_{=}(p) $ is defined
by,
\begin{eqnarray}
0 
& = & 
k(p)^2 - (\frac{1}{1 + \kappa_{=}} k(p) - \frac{1}{2} \frac{\alpha}{1 + \alpha \kappa_{=}}
(1 + p^2)) k(p) 
\nonumber \\
&   &
- \frac{1}{1 + \kappa_{=}} \frac{\alpha}{1 + \alpha \kappa_{=}} p
\end{eqnarray}
which after some manipulation may be re-arranged to give,
\begin{equation}
0 = \kappa_{=}^2 + B(p, \alpha) \kappa_{=} - C(p, \alpha)
\end{equation}
where $ B(p, \alpha) \equiv 1 + 1/\alpha + (1 - p)/k(p) $,
$ C(p, \alpha) \equiv \frac{1}{2} \frac{1}{k(p)}(1 + 1/k(p)) (1 - p^2) $.
We then have that,
\begin{equation}
\kappa_{=}(p) = B (\sqrt{1 + 4 \frac{C}{B^2}} - 1)
\end{equation}

We claim that $ \kappa_{=}(p) $ is a decreasing function
of $ p $ for $ p \in [p_{crit}, 1] $.  We will prove this
by showing that $ B $ and $ C/B^2 $ are both decreasing functions
of $ p $ for $ p \in [p_{crit}, 1] $.  

To prove that $ B $ is a decreasing function of $ p $ for
$ p \in [p_{crit}, 1] $, we need show that $ (1 - p)/k(p) $
is a decreasing function of $ p $ for $ p \in [p_{crit}, 1] $.
Differentiating, we obtain,
\begin{equation}
\frac{d}{d p} (\frac{1 - p}{k(p)}) = -\frac{1}{2} \frac{1 - p^2 + 2 p}{k(p)^2} < 0
\end{equation}
so $ B $ is certainly a decreasing function of $ p $ for $ p \in [p_{crit}, 1] $.

Now, after some manipulation, we can show that,
\begin{equation}
\frac{C}{B^2} = \frac{1}{2} \alpha^2 \frac{1 - p^2}
{(1 + k(p)) (1 + \alpha - \frac{1 + \alpha p}{1 + k(p)})^2}
\end{equation}

Note that $ 1 - p^2 $ is decreasing for $ p \in [p_{crit}, 1] $,
and that $ 1 + k(p) $ is increasing.  We also have that,
\begin{equation}
\frac{d}{dp} (\frac{1 + \alpha p}{1 + k(p)})
= -\frac{(\alpha/2) p^2 + p + (1 - \alpha/2)}{(1 + k(p))^2} < 0
\end{equation}
so that $ (1 + \alpha p)/(1 + k(p)) $ is a decreasing function
of $ p $.  Therefore, $ 1 + \alpha - (1 + \alpha p)/(1 + k(p)) $
is an increasing function of $ p $, hence
$ C/B^2 $ is decreasing for $ p \in [p_{crit}, 1] $, as we wished
to show.

Figure 3 illustrates the behavior of $ \kappa_{=}(p) $ for three
values of $ \alpha $.

\begin{figure}
\includegraphics[width = 0.9\linewidth, angle = 0]{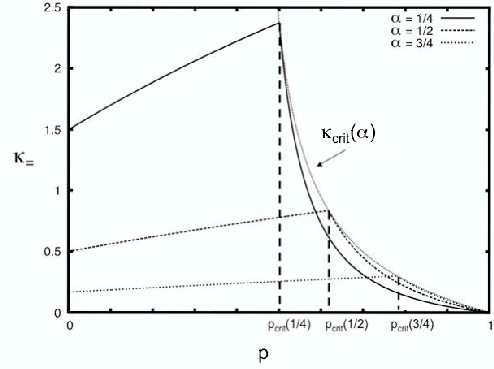}
\caption{$ \kappa_{=} $ versus $ p $ for $ \alpha = 1/4, 1/2, 3/4 $.
A graph of $ \kappa_{crit}(\alpha) $ is included as well.}
\end{figure}

In what follows, we shall change our notation slightly to 
explicitly indicate that $ \kappa_{=} $ also depends on
$ \alpha $.  Thus, we shall re-denote $ \kappa_{=}(p) $
by $ \kappa_{=}(p, \alpha) $.  This notation was not
needed in the previous arguments, since we were considering
the behavior of $ \kappa_{=} $ at a fixed $ \alpha $.

We may now summarize the behavior of $ \kappa_{=}(p, \alpha) $
as a function of $ p $:

From $ 0 $ to $ p_{crit} $, $ \kappa_{=}(p, \alpha) $ increases
from $ \frac{1}{2} (1 - \alpha)/\alpha $ to $ \kappa_{crit}(\alpha) \equiv
\kappa_{=}(p_{crit}(\alpha), \alpha) $, while from $ p_{crit} $ to $ 1 $, 
$ \kappa_{=}(p, \alpha) $ decreases from $ \kappa_{crit}(\alpha) $ to $ 0 $.  This behavior leads to
three distinct regimes of $ \kappa $.

For $ \kappa \in [0, \frac{1}{2} (1 - \alpha)/\alpha] $, there exists
only one value of $ p $ for which asexual and sexual replication
yield identical mean fitness results.  This value of $ p $ is
contained in the interval $ [p_{crit}, 1] $.  As $ \kappa $ 
increases from $ 0 $ to $ \frac{1}{2} (1 - \alpha)/\alpha $, this
value of $ p $ decreases.  For these values of $ \kappa $, 
asexual replication is advantageous over sexual replication at
low mutation rates.  However, there is a crossover replication
fidelity where sexual replication becomes advantageous.  As 
$ \kappa $ increases, this crossover replication fidelity gets
pushed to lower values.  This makes sense, since a higher
value of $ \kappa $ corresponds to a greater penalty associated
with sex.

For $ \kappa \in [\frac{1}{2} (1 - \alpha)/\alpha, \kappa_{crit}(\alpha)] $,
there exist exactly two values of $ p $ for which asexual and sexual replication
yield identical mean fitness results.  One value of $ p $ is contained in
the interval $ [0, p_{crit}] $, while the other value is contained
in the interval $ [p_{crit}, 1] $.  As $ \kappa $ increases from
$ \frac{1}{2} (1 - \alpha)/\alpha $ to $ \kappa_{crit}(\alpha) $, the
value of $ p $ in $ [0, p_{crit}] $ increases from $ 0 $ to $ p_{crit}(\alpha) $,
while the value of $ p $ in $ [p_{crit}, 1] $ decreases to $ p_{crit}(\alpha) $.
For these values of $ \kappa $, asexual replication is also advantageous over sexual
replication at low mutation rates.  As with the previous regime, there is a crossover
replication fidelity where sexual replication becomes advantageous.  However, in
contrast to the first $ \kappa $ regime, there is a second crossover replication
fidelity where asexual replication again becomes advantageous.  For these values of
$ \kappa $, the cost associated with sex is still sufficiently low that sexual
replication can become the advantageous strategy at higher mutation rates.  However,
the cost of sex is sufficiently high that, at even higher mutation rates, sexual
recombination no longer offsets the production of unviable chromosomes from viable ones
to an extent that makes the strategy advantageous.

Finally, for $ \kappa \in (\kappa_{crit}(\alpha), \infty) $, the cost 
associated with sex is so high that sexual replication is never the advantageous
strategy.

As a final note for this subsection, we can show that $ \kappa_{crit}(\alpha) $
is a decreasing function of $ \alpha $.  Differentiating the quadratic equation given by
Eq. (19) with respect to $ \alpha $ (where $ x = x(\alpha) \equiv \alpha \kappa_{crit} (\alpha) $), we
have,
\begin{eqnarray}
0 
& = & 
(2 x p_{crit} + \alpha p_{crit} + 1) \frac{d x}{d \alpha} 
\nonumber \\
&   &
(x^2 + \alpha x - (1 - \alpha) (1 + p_{crit})) \frac{d p_{crit}}{d \alpha}
\nonumber \\
&   &
+ p_{crit} x + \frac{1}{2} (1 + p_{crit})^2
\end{eqnarray}
Differentiating both sides of the equality defining $ p_{crit} $, it is possible
to show that $ d p_{crit}/d \alpha = p_{crit}/(1 - \alpha + p_{crit}) $.
We also have, from a previous analysis, 
that $ p x^2 + \alpha x p - (1 - \alpha) p (1 + p) = 
\frac{1}{2} (1 - \alpha) (1 - p^2) - x $, so that, to show
$ d x/d \alpha < 0 $, we need to show that
\begin{equation}
0 < \frac{\frac{1}{2} (1 - \alpha)(1 - p_{crit}^2) - x}{1 - \alpha + p_{crit}} +
p_{crit} x + \frac{1}{2} (1 + p_{crit})^2 
\end{equation}
Multiplying by $ 1 - \alpha + p_{crit} $ and simplifying, this is equivalent to the
inequality,
\begin{equation}
0 < (1 - \alpha - x) + p_{crit} [(1 - \alpha) (1 + x) + x p_{crit} + \frac{1}{2} (1 + p_{crit})^2]
\end{equation}
Since we showed that the expression for $ x(p) $ in Eq. (19), valid over $ p \in [0, p_{crit}] $, is an increasing
function of $ p $ for $ p \in [0, 1] $, then solving the quadratic in Eq. (19) for $ p = 1 $ gives
$ x \leq 1 - \alpha $.  Therefore, $ 1 - \alpha - x \geq 0 $, hence the inequality holds.

We have therefore shown that $ \alpha \kappa_{crit}(\alpha) $ is a decreasing function
of $ \alpha $, hence $ \kappa_{crit}(\alpha) $ is a decreasing function of $ \alpha $.
When $ \alpha = 1 $, $ p_{crit} = 1 $, so $ \kappa_{=} = 0 $.  As $ \alpha \rightarrow 0 $,
$ p_{crit} \rightarrow \sqrt{2} - 1 $, so $ \alpha \kappa_{=} \rightarrow
(\sqrt{4 \sqrt{2} - 3} - 1)/[2 (\sqrt{2} - 1)] \Rightarrow \kappa_{=} \rightarrow \infty $.

\subsection{Consideration of $ \kappa_{UU} > 0 $}

When $ \kappa_{UU} > 0 $, the results for sexual replication
remain unchanged.  However, the results for asexual replication
change somewhat, since now an additional localization to
delocalization transition can occur, once $ \kappa_{UU} =
\max \{\kappa_{VV} (2 ((1 + p)^2/2) - 1), \kappa_{VU} p,
\kappa_{UU} \} $.  It is therefore possible to have a situation
where sexual replication only becomes advantageous once
the mean fitness of an asexually replicating population
is $ \kappa_{UU} $.  With an appropriate choice of parameters,
it is then possible that the mean fitness of the sexually
replicating population is less than $ \kappa_{UU} $, so that
sexual replication will never be the preferred replication
strategy.  We leave the investigation of this phenomenon for
future work.

In any event, for $ \kappa_{UU} > 0 $, asexual replication will 
become the advantageous mode of replication at sufficiently
high mutation rates, since the mean fitness of the sexually
replicating population decreases to zero, while after complete
delocalization over the genome space has occurred, the mean
fitness of the asexually replicating population becomes $ \kappa_{UU} $.
This result, however, is likely due to a mating strategy that
essentially ``throws away'' the ``U'' chromosomes.  Other mating
strategies, where all haploids are capable of mating, may not
exhibit the same effect.

\section{Discussion, Conclusions, and Future Research Directions}

This paper developed a simplified model for sexual replication, and showed,
within the context of the model, that a sexually replicating population will
outcompete an asexually replicating one when there there is no cost associated
with sex.  We further showed that if the cost associated with sex is sufficiently
low, then sexual replication becomes advantageous at higher mutation rates, 
because recombination prevents the accumulation of defective mutations in the
diploid genomes (assuming that there is a fitness penalty associated with the
defective mutations).  The cost for sex was measured by the dimensionless parameter
$ \kappa $, defined to be the product of the first-order growth rate constant of
the mutation-free genomes ($ \kappa_{VV} $), and the characteristic time associated
with finding a recombination partner ($ \tau_{seek} $).  Since $ \kappa_{VV} =
1/\tau_{rep} $, where $ \tau_{rep} $ denotes a characteristic replication time,
it follows that $ \kappa = \tau_{seek}/\tau_{rep} $.  Therefore, the cost associated
with sex is measured by the ratio of the time a haploid spends finding a recombination
partner with the time scale for replication.  The smaller this ratio, the smaller
the fitness penalty incurred by reproducing via a haploid intermediate, and the
greater the selective advantage for sex.

The implications of this model are that sexual replication is favored in environments
where organisms replicate relatively slowly, and in environments where the time
spent finding a recombination partner is small compared with the time scale for
replication.  Thus, sexual replication is favored in environments with high population
density.  These results are therefore consistent with the observation that sexual
replication is the preferred (and generally the only) mode of reproduction for nearly
all multicellular organisms.  

That sexual replication only becomes the preferred mode of reproduction at low $ \kappa $
suggests why sexual replication occurs as a stress response in some organisms,
such as {\it Saccharomyces cerevisiae} (Baker's yeast).  When conditions are favorable,
$ \kappa_{VV} $, and hence $ \kappa $, are relatively high, so asexual replication
is the advantageous strategy.  Under sufficiently adverse conditions, $ \kappa $ can
drop to levels where the sexual strategy becomes advantageous.  The replicative
strategy that can adopt the optimal replication strategy for the given environment
will have a selective advantage (assuming that resource costs for maintaining
this switching behavior are not prohibitive), and so organisms carrying this strategy
in their genomes will dominate the population.

However, as one moves toward more complex life forms, the replication rate
drops to values such that asexual replication is almost never the preferred
mode of reproduction, so that the ability for an organism to switch between
the two modes of reproduction disappears.  At this point, we postulate that
the division of haploid cells into two distinct types of gametes, and then 
later the division of the organisms themselves into male and female, are the
result of selection for evolutionary pathways leading to the division
of labor and specialization of tasks associated with sexual replication.
When replication rates are low, and when the time cost associated with sex
is low, then it is likely more efficient (in terms of resource utilization) 
to divide the reproductive tasks associated with sexual replication among two 
types of organisms (``male'' and ``female'').  The relative fitness advantage
as a result of such savings in resource costs likely increases with the complexity
of the organism, leading to a stronger selection pressure for a male-female
split as organismal complexity grows.

In this paper, we assumed that only haploid cells with viable chromosomes are 
capable of engaging in sexual recombination.  This allowed a simplified analysis 
within the standard quasispecies formalism.  While we obtained a selective advantage for sexual
replication using this mating strategy, a fuller analysis will
require the consideration of various mating strategies on the
selective advantage for sex.  An important such mating strategy,
which is the opposite of the one considered in this paper,
is the random mating strategy, whereby all haploids are capable of
engaging in sexual recombination, and do so with a pairwise distribution 
given by the Hardy-Weinberg equilibrium.

In this vein, one interesting question is to determine, for a given fitness landscape, 
whether there always exists a mating strategy for which sexual replication will outcompete asexual
replication.  Additionally, while this paper implicitly assumed that the strategy for
sexual or asexual replication is inherited, future studies should consider genomes where
genes for sex are explicitly included.  This leads to the ability for sexual organisms
to mutate into asexual ones.  As the selective advantage for sexual replication
disappears (as a function of mutation rate, for instance), the models may exhibit
localization to delocalization transitions over the portions of the genome controlling
sex.

\begin{acknowledgments}

This research was supported by the Israel Science Foundation.

\end{acknowledgments}


\begin{thebibliography}{50}
\expandafter\ifx\csname
natexlab\endcsname\relax\def\natexlab#1{#1}\fi
\expandafter\ifx\csname bibnamefont\endcsname\relax
  \def\bibnamefont#1{#1}\fi
\expandafter\ifx\csname bibfnamefont\endcsname\relax
  \def\bibfnamefont#1{#1}\fi
\expandafter\ifx\csname citenamefont\endcsname\relax
  \def\citenamefont#1{#1}\fi
\expandafter\ifx\csname url\endcsname\relax
  \def\url#1{\texttt{#1}}\fi
\expandafter\ifx\csname
urlprefix\endcsname\relax\def\urlprefix{URL }\fi
\providecommand{\bibinfo}[2]{#2}
\providecommand{\eprint}[2][]{\url{#2}}

\bibitem[{\citenamefont{Stauffer et~al.}(1997)}]{SEX1}
\bibinfo{author}{\bibfnamefont{D.}~\bibnamefont{Stauffer}}, 
\bibinfo{author}{\bibfnamefont{P.M.C.}~\bibnamefont{de Olveira}},
\bibinfo{author}{\bibfnamefont{S.}~\bibnamefont{Moss de Olveira}},
\bibnamefont{and}
  \bibinfo{author}{\bibfnamefont{R.M.}~\bibnamefont{Zorzenon}},
  \bibinfo{journal}{cond-mat/9605110}, (\bibinfo{year}{1997}).

\bibitem[{\citenamefont{Sa Martins and Stauffer}(2001)}]{SEX2}
\bibinfo{author}{\bibfnamefont{J.S.}~\bibnamefont{Sa Martins}} \bibnamefont{and}
  \bibinfo{author}{\bibfnamefont{D.}~\bibnamefont{Stauffer}},
  \bibinfo{journal}{cond-mat/0102176}, (\bibinfo{year}{2001}).

\bibitem[{\citenamefont{He et~al.}(2003)}]{SEX3}
\bibinfo{author}{\bibfnamefont{M.}~\bibnamefont{He}}, 
\bibinfo{author}{\bibfnamefont{H.}~\bibnamefont{Ruan}},
\bibinfo{author}{\bibfnamefont{C.}~\bibnamefont{Yu}}
\bibnamefont{and}
  \bibinfo{author}{\bibfnamefont{L.}~\bibnamefont{Yao}},
  \bibinfo{journal}{cond-mat/0309187}, (\bibinfo{year}{2003}).

\bibitem[{\citenamefont{Holmstrom and Jensen}(2003)}]{SEX4}
\bibinfo{author}{\bibfnamefont{K.}~\bibnamefont{Holmstrom}} \bibnamefont{and}
  \bibinfo{author}{\bibfnamefont{H.J.}~\bibnamefont{Jensen}},
  \bibinfo{journal}{cond-mat/0309300}, (\bibinfo{year}{2003}).

\bibitem[{\citenamefont{Sa Martins and Racco}(2001)}]{SEX5}
\bibinfo{author}{\bibfnamefont{J.S.}~\bibnamefont{Sa Martins}} \bibnamefont{and}
  \bibinfo{author}{\bibfnamefont{A.}~\bibnamefont{Racco}},
  \bibinfo{journal}{cond-mat/0011499}, (\bibinfo{year}{2001}).

\bibitem[{\citenamefont{Eigen}(1971)}]{EIG}
\bibinfo{author}{\bibfnamefont{M.}~\bibnamefont{Eigen}},
  \bibinfo{journal}{Naturewissenschaften} \textbf{\bibinfo{volume}{58}},
  \bibinfo{pages}{465} (\bibinfo{year}{1971}).

\bibitem[{\citenamefont{Alves and Fontanari}(1997)}]{DIP}
\bibinfo{author}{\bibfnamefont{D.}~\bibnamefont{Alves}} \bibnamefont{and}
  \bibinfo{author}{\bibfnamefont{J.F.}~\bibnamefont{Fontanari}},
  \bibinfo{journal}{J. Phys. A:  Math Gen.} \textbf{\bibinfo{volume}{30}},
  \bibinfo{pages}{2601} (\bibinfo{year}{1997}).

\end{thebibliography}
\end{document}